# AI Personalization Paradox: Personalized AI Increases Superficial Engagement in Reading while Undermines Autonomy and Ownership in Writing

Peinuan Qin
School of Computing, National
University of Singapore
Singapore, Singapore
e1322754@u.nus.edu

Chi-Lan Yang
Graduate School of Interdisciplinary
Information Studies, The University
of Tokyo
Tokyo, Japan
chilan.yang@cyber.t.u-tokyo.ac.jp

Nattapat Boonprakong
School of Computing, National
University of Singapore
Singapore, Singapore
nattapat.boonprakong@nus.edu.sg

Jingzhu Chen
School of Computer Science and
Technology, Tongji University
Shanghai, China
2253543@tongji.edu.cn

Yugin Tan
Computer Science, National
University of Singapore
Singapore, Singapore
tan.yugin@u.nus.edu

Yi-Chieh Lee
National University of Singapore
Singapore, Singapore
yclee@nus.edu.sg

## Abstract
AI-assisted writing raises concerns about autonomy and ownership when benefiting writers. Personalization has been proposed as an effective solution while also risking writers' reliance on AI and behavior shifting. For better personalization design, existing studies rely on interaction and information solely within the writing phase; however, few studies have examined how reading behaviors can inform personalized writing. This study investigates the effects of integrating reading highlights for personalization on AI-assisted writing. A between-subjects study with 46 participants revealed that the personalization condition encouraged participants to produce more highlights. However, highlighting unexpectedly shifted from a sense-making strategy to an instrumental act of "feeding the AI," leading to significant reliance on AI and declines in writers' sense of autonomy, ownership, and self-credit. These findings indicate personalization risks in AI-assisted writing, emphasize the importance of personalization strategies, and provide design implications.

## 1 Introduction

AI-assisted writing tools have been increasingly adopted in various applications, including argumentative writing [48, 81], essay composition [31], and storytelling [77, 95]. However, despite their benefits in improving writing [63], research has raised concerns that AI-assisted writing can harm user autonomy and their perceived ownership [8, 83]. Qin et al. [62] demonstrated that AI assistance in writing ideation reduced ideation quantity, as well as the writers' ownership, autonomy, and self-credit when there was no independent ideation of writers.

To mitigate these issues, many efforts have focused on *personalization*, which tailors AI support to an individual's needs, preferences, or characteristics [44, 50, 88]. Personalization has been widely explored in education [44] and writing [93] as a means to restore control. For example, GhostWriter [93] enables writers to guide AI's tone, fostering closer alignment, sense of ownership, and satisfaction with AI-generated suggestions.Personalization has also been shown to promote more active engagement in learning [21].

On the other hand, personalization can also backfire, exacerbating users' over-reliance on AI [34] and leading to unconscious shifts in individual behaviors. Implicit egotism [60] suggests that people unconsciously prefer outputs linked to themselves. This indicates that generating personalized suggestions aligned with users' values may lead to an illusory reliance on AI instead of encouraging users to engage critically with the task [28, 34, 74]. Other studies [6, 11, 56] further indicated that personalization can nudge users to shift their behaviors to overly adapt to the system. Such adaptation can limit users' creativity, critical thinking, and awareness of alternative viewpoints, which are important cognitive processes in writing. These findings highlight the need for further exploration to understand the risks associated with personalization in AI-assisted writing, which will inform the design of such personalized systems.

However, most existing work on personalized AI-assisted writing has concentrated narrowly on the writing stage [2, 91, 93]. GhostWriter [93], for example, captured authors' style through writing interactions. Tang et al. [82] personalized the AI's writing assistance by summarizing users' writing history into personal profiles; nonetheless, they also relied on the information in the writing stage. While these systems demonstrate promise, writing is rarely an isolated act. Reading and writing are deeply intertwined, sharing cognitive processes and knowledge [32, 45, 73]. Especially in source-based writing tasks [27, 33, 53], such as argumentative writing, writers rely heavily on reading to understand texts, accumulate materials and evidence for subsequent synthesis and argument development. In reading, annotations such as highlights are commonly used to deepen comprehension and distinguish important content [37, 52, 96]. Moreover, it can externalize one's cognitive process [64, 71], representing their reading focus [90, 94], suggesting its potential to serve as an important medium for conveying writers' writing intentions to AI. However, current research has not fully explored the role of including reading highlights in promoting personalized AI-assisted writing.

To address this gap, in this paper, we consider highlights during the reading process as a means to achieve personalization in AI-assisted writing. By highlighting important content, users can finely control the context range used for generating AI writing suggestions. Specifically, we investigate how this personalization approach would influence users' behaviors and perceptions in both reading and writing. Specifically, we ask:

**RQ1:** How does using highlights to personalize AI-assisted writing influence users' reading strategies?

**RQ2:** How does using highlights to personalize AI-assisted writing influence users' behaviors and perceptions in writing?

**RQ3:** How does highlight-based AI personalization change the relationship between reading and writing?

We conducted a between-subjects experiment with 46 participants, who were tasked with writing an argumentative essay based on given reading materials. Concretely, participants first completed a reading phase in which they could freely highlight information they considered important. They then moved to the writing phase, during which they could request AI-generated suggestions. In *Personalization* condition, AI suggestions were generated solely based on participants' highlights in reading materials, while in *Baseline*, the AI suggestions were based on full reading materials.

The results showed that when using reading highlights for personalized AI writing assistance, the participants engaged more in reading, with more highlights and longer reading durations. However, this engagement was superficial and did not contribute to participants' positive writing experience. Instead, their writing behaviors featured fewer edits and lower revision intensity, indicating a higher reliance on AI, alongside reduced autonomy, ownership, and credit attribution. This unexpected finding suggests a shift in motivation that reading engagement became extrinsically oriented toward producing inputs for AI suggestions rather than intrinsically deepening understanding. In contrast, in the *Baseline* condition, effortful reading and active highlighting behaviors correlate with positive writing behaviors and perceptions, including greater perceived effort, more active editing, stronger autonomy, enhanced ownership, and increased self-credit. Taken together, our results suggest the need for effective designs that integrate the reading process with personalized AI-assisted writing to promote better writing experience.

The contribution of this study is two-fold:

(1) We empirically bridge the gap, integrating reading for personalized AI-assisted writing tools. Much of the prior HCI research on AI writing assistants has focused on the writing stage in isolation. This study provides the first empirical evidence on how interactions in the reading phase directly impact behaviors and perceptions in the writing phase.

(2) We identified a paradoxical phenomenon in which a seemingly helpful personalization design actually promotes superficial engagement with writing and leads to negative writing perceptions. This phenomenon provides a valuable example for discussing a negative interaction pattern that designers should consider when designing personalized AI-assisted tools.

## 2 Related Work

### 2.1 Development and Challenges of AI-Assisted Writing

With the development of large language models (LLMs), AI-assisted writing tools have emerged to enhance writing processes and boost productivity. Reza et al. [63] surveyed existing AI-assisted writing systems and suggested that these systems are typically designed to support three stages of writing, as previously identified by Flower and Hayes [23]: planning, translation, and revision. For example, Shaer et al. [72] targeted the planning stage and developed a collaborative group-AI ideation framework that integrated an LLM to support idea generation and solution exploration. Bhat et al. [7] focused on the translation stage by implementing next-phrase suggestion tools to guide writers' ongoing composition, while Dhillon et al. [18] explored how varying levels of scaffolding, at the sentence or paragraph level, shape the co-writing process with AI. For the revision stage, Liu et al. [49] designed *eRevise+RF* as an AI assistant to evaluate and guide student essay revisions.

Despite these benefits, growing evidence shows that AI-assisted writing systems also raise concerns, especially about user autonomy [22, 25], ownership [83], credit attribution [62], and reliance on AI [74]. For instance, Draxler et al. [20] found that writers often struggled to claim ownership of AI-generated text, even when writers presented it as their own. Others observed that writers tended to attribute more credit to the AI than to themselves [62, 87], especially when AI was introduced from the outset [62]. These research findings collectively suggest the need to explore a way to balance writers' autonomy with the supportive capabilities that AI can offer.

### 2.2 Personalization in AI-Assisted Writing

To enable people's autonomy while using AI-assisted tools, personalization is a common approach [20, 34, 84, 93]. Personalization refers to the process of tailoring AI support to an individual's specific needs, preferences, or characteristics, often by leveraging users' prior interactions, styles, or behaviors [44, 50, 88]. While personalization is often praised for enhancing autonomy, ownership, and engagement, it can also lead to over-reliance, reduce critical thinking, and nudge users to overly adapt their behaviors to the system [6, 56]. The personalization process is in its early stages, primarily concentrating on achieving personalization during the writing interactions [82, 93] and not adequately exploring the reading phase. Notably, the literature suggests that reading highlights [52, 64, 71, 90] carries substantial potential to enhance personalized AI-assisted writing. Thus, this section aims to clarify the rationale for integrating highlights in personalized AI-assisted writing design.

*2.2.1 Benefits of Personalized AI-Assisted Writing.* Existing research suggests personalization has the potential to enhance autonomy and ownership [20, 34, 84, 93]. As an example of personalized AI systems for writing, GhostWriter [93] learns an author's writing style implicitly while offering explicit style control via manual annotations, which enhances personalization and user agency, helping authors craft text that aligns with their style preferences. In the context of education, prior studies have already demonstrated how personalization improved students' engagement [38, 42, 92]. Effort justification [19, 51] and the IKEA effect [55, 70] further indicate the importance of effort investment in establishing ownership. These findings suggest that writers in personalized AI-assisted writing may also become more engaged in the writing process and sustain a stronger sense of ownership.

*2.2.2 Risks of Personalized AI-Assisted Writing.* However, personalization can introduce harm to users. Bellomo [6] distinguished different personalization types in an education context: extrinsic personalization, which involves aligning external systems with users' behavior, and intrinsic personalization, which focuses on cultivating learners' autonomous capabilities for reflection and meaning-making, and argued that extrinsic personalization risks sustaining passivity and constraining students to algorithmically predetermined paths [14]. Moreover, implicit egotism [60] suggests that humans have an unconscious preference for items they associate with themselves. In the context of AI-assisted writing, studies found that writers tended to favor personalized suggestions from AI, thus causing them to overrely on AI and diminishing their critical thinking [28, 74]. Similarly, Hwang et al. found that while writers generally preferred personalization, they also perceived it as a double-edged sword: while personalization may preserve writers' authentic writing voice, it may exacerbate overreliance on AI [34].

Additionally, Ontanon and Zhu [56] suggested that personalization can push users to shift their behavior to adapt to the system. The authors coined the "personalization paradox": *"there lies a more general paradox at the very heart of personalization. Personalization promises to modify your digital experience based on your personal interests and preferences. Simultaneously, personalization is used to shape you, to influence you, and guide your everyday choices and actions."* Similar examples are widely found in recommendation systems [12, 13, 30]. Hannak et al. [30] found that users on online platforms adjusted their behavior to improve their visibility in search rankings, such as timing their posts strategically. In music recommendation, nearly 50% of participants reported that they consciously altered their behavior in real-world platforms to influence future recommendations [12]. This adaptation can possibly limit authors' creativity, critical thinking, and awareness of alternative viewpoints [6, 56].

*2.2.3 Limitations of Existing Personalized AI-Assisted Writing.* Existing research in AI-assisted writing suggests that the design paradigm is fixated primarily on the writing stage [2, 93]. For example, GhostWriter [93] captures a writer's writing style implicitly while offering explicit style control via manual annotations. These behaviors of teaching LLM rely solely on the writing stage. Besides, InkWell [24] serves as a creative writer's assistant that focuses on stylistic variation and writer mimicry. In addition, Tang et al. [82] personalizes the LLM writing assistant by distilling users' scientific writing history (i.e., essential traits and preferences) into concise personal profiles [82]. In these works, the information, based on which to implement personalization [24, 82], is largely limited to the writing stage. Such personalization approaches do not accommodate source-based writing processes, such as argumentative writing, where users rely on cognitive foundations from reading [45], switch frequently between reading and writing, and build their own knowledge, arguments, and mark evidence through annotation [27, 53]. Therefore, the reading process should be taken into consideration when designing personalized AI-assisted writing systems, rather than viewing writing as an isolated task.

## 2.3 Potential of Highlights for Personalized AI-Assisted Writing

Reading and writing are closely intertwined both cognitively and behaviorally, rather than being isolated processes. Prior work highlights that they share underlying cognitive mechanisms such as domain knowledge, procedural knowledge, and metacognitive strategies, enabling readers and writers to engage in meaning-making activities [32, 45]. Behaviorally, writing often involves iterative switching with reading, particularly in source-based writing tasks where writers must organize and synthesize information from multiple sources while ensuring appropriate citation and integration [27, 53]. These findings suggest that reading behaviors can provide valuable signals for supporting writing.

Among reading behaviors, highlighting is commonly used to facilitate comprehension and memory by filtering redundant details, thereby reducing cognitive load and supporting smoother integration of evidence into writing [37, 65]. In addition, highlights can stand out as a salient externalization of user cognition [64, 71], they represent individuals' cognitive priorities and interests in a tangible form. These properties underscore a promising signal for personalization, offering AI systems insight into what matters most to writers and enabling more tailored and supportive writing assistance. However, highlights have not been given attention in the current personalized writing solutions.

To address this gap, our study leverages reading highlights for a personalized AI-assisted writing experience and explores how this personalization process affects the reading and writing stages of writers.

## 3 Method
## 3.1 Study Design

We conducted a between-subjects study with two experimental conditions that manipulated whether participants' highlights were used to generate personalized AI suggestions. We refer to these conditions as *Personalization* and *Baseline*.

*Condition Settings.* To ensure an accurate understanding of the system's behavior, participants in each condition were explicitly informed about how the AI would use the reading materials. In the *Personalization* condition, participants were told that the AI system would generate suggestions *solely* from the sentences or phrases they highlighted during reading. In the *Baseline* condition, participants were informed that the AI would instead rely on the *full set* of source texts, regardless of their individual highlights.

*Rationale for the Condition Design.* A feasible alternative is to use full-text as the base configuration in the *Personalization* condition and build a personalized generated context through highlights on this basis. However, if we used this approach, we cannot intentionally engage participants in highlighting for getting personalized suggestions in *Personalization* condition because participants could potentially not engage in using any highlights to obtain AI-assisted writing, which may make such manipulation invalid. Prior work on default effects [89] and status quo bias [68] showed that users tend to stay with system-provided defaults rather than investing effort to reconfigure settings or explore alternative modes of use to reduce mental effort and friction. Thus, we set the default option in the *Personalization* condition to "AI writing suggestion based on highlights," ensuring our manipulation is valid.

*Writing Task.* The writing task for participants was an *argumentative essay* [53, 63], which required authors to take a clear stance on a controversial issue, support their position with evidence and reasoning to persuade the reader. We chose the argumentative essay as a writing task because it is a source-based writing.

## 3.2 Materials and Task Arrangement

To ensure topic diversity and relevance, we selected four debate-style topics commonly used in argumentative writing: *social media*, *assisted suicide*, *college education*, and *vaping*. In addition, we also selected an additional *school uniform* topic for the practice task. Each topic included two source articles that take opposing stances (e.g., on the social media topic, one article argues that social media is good for society, while another article suggests that social media is bad for society). Each of the source articles is written in English, with a word count of approximately 1100-1200. We sourced the articles from the New York Times. To eliminate the interference of format with content, we manually process these articles into simple text files. The articles used in this study are included in the supplementary materials.

To prevent participants from being influenced by their familiarity or interest in the topic when completing writing tasks, we screened participants through questions in the pre-survey. For control, we embedded the description of the 4 topics in the pre-survey. Each participant needed to fill out a 7-point scale (1 = Strongly Disagree; 7 = Strongly Agree) regarding their familiarity and interests in the different topics [36]; for example, in the topic of social media, our items were: *"I feel the issue of Social Media is related to my core value"*, *"I feel it is important to defend my point of view on the issue of Social Media"*, *"I feel I am interested in learning about the issue of Social Media"*, *"I feel I desire to know the facts about the issue of Social Media"*, and *"I am familiar with the issue of Social Media."* The averaged score was used for the final selection.

Previous studies have shown that participants' interest and familiarity [3, 16, 39] influence writing behaviors and performance; therefore, we adopt these two indicators as criteria to determine whether participants are suitable for writing on a specific topic. We collected data from all registered participants and calculated each person's scores on different topics. For each topic, we used the mean and one standard deviation range ($M \pm SD$) as a standard. Participants whose topic score will be considered as a candidate on this topic, and each candidate may be suitable for multiple topics simultaneously. We will ultimately dynamically allocate participants to the topic based on the number of people in each topic.

We conducted a two-way ANOVA on participants' familiarity and interest scores to examine whether there were differences across conditions and topics. The analysis revealed no significant main effect of condition, $F(1, 38) = 0.027$, $p = 0.871$, indicating that participants in different experimental conditions did not differ in their topic familiarity and interest. There was also no significant main effect of topic, $F(3, 38) = 1.841$, $p = 0.156$, suggesting that the four topics were balanced in terms of participants' reported familiarity and interest. Finally, the condition × topic interaction was not significant, $F(3, 38) = 0.889$, $p = 0.456$. These results confirm that our topic allocation procedure successfully balanced participants across conditions and topics, eliminating potential confounds from prior familiarity or interest.

## 3.3 Participants

Participants (N = 50) were recruited using snowball sampling from the local community through online advertisements posted in public social media channels and community groups. No distinction was made regarding participants' occupation or academic background. Therefore, we reached out to the local network of university students, staff, and affiliates. Inclusion criteria required that participants were native English readers and had prior experience with argumentative writing. Participants were randomly assigned to one of the two experimental conditions. After the experiment, data from four participants were discarded due to failing the manipulation check (Section 3.6.1), leaving 46 participants for data analysis–22 in the *Baseline* condition and 24 in the *Personalization* condition. Among the included participants, 27 identified as female (59%) and 19 indicated as male (41%). The age range was between 21 to 54 years (M = 28.957; SD = 7.726). Regarding educational level, the majority of participants (85%) indicated having obtained at least a high school degree. All participants reported they had experience with argumentative writing. Each participant received a compensation of USD $13 (local equivalent) for a 90-minute study upon completion. The study was approved by the Institutional Review Board (IRB) of the National University of Singapore (NUS).

## 3.4 Procedure

This study was conducted remotely via Zoom. In each session, an experimenter provided a brief overview of the study and assisted participants in setting up screen sharing for recording purposes. Participants were explicitly informed that screen sharing was used solely to retain evidence of task progress and not for real-time monitoring. The experimenter turned off their camera and left the shared screen view once the study began.

The entire study lasted approximately 90 minutes, and participants followed the steps. **(A) *Introduction and Warm-Up.*** Participants read a brief study overview and provided informed consent. Before starting the experiment, the experimenter showed participants a brief video tutorial about how to use our system. Then, participants went through a warm-up reading-writing task to familiarize themselves with our protocols. They engaged in reading and writing on a topic different from the actual task, with a shorter reading length and a brief argumentative writing (100 - 200 words). The introduction and warm-up stage took approximately 15 minutes. **(B) *Reading Task.*** Participants were instructed to read the articles, in which they could navigate within or between articles and highlight information. Participants notified the experimenter when they were ready to proceed to writing. **(C) *Writing Task.*** Participants were asked to write an argumentative essay (350 - 500 words) based on information from the source articles. During this phase, AI suggestions were generated upon participants' request and shown according to the assigned condition. During writing, they were allowed to switch between the reading and writing interfaces. **(D) *Post-task Survey.*** After the reading and writing task, participants completed a questionnaire reporting their perceptions during the AI-assisted writing process.

## 3.5 Experiment Platform

*3.5.1 System Overview.* We developed a custom web-based platform. The platform guides participants through a structured reading-and-writing workflow: they first read source documents and highlight important content, then compose an argumentative essay while receiving AI-generated suggestions. The system was built with a Vue.js[1] front end, a Django[2] back end, deployed on AWS EC2[3] for stable access. We used OpenAI's GPT-4.1 API[4] for writing suggestion generation. To generate proper AI suggestions, the system followed the retrieval-augmented generation (RAG) workflow [26, 67], separately storing the split paragraphs of the reading content and the highlighted content in a vector database, Chroma[5], using OpenAI's *text-embedding-3-large* model for embedding. We applied a reranking step using Cohere's *rerank-v3.5* model to prioritize the most relevant content before generating AI suggestions. The technical architecture diagram is shown in Fig. 1. During the reading phase (A), in the *Baseline* condition, when a document is input into the system, it is divided into fixed-size text chunks, with each chunk having a limit of 300 tokens and maintaining an overlap of 50 tokens to preserve context. These chunks provide the raw material for the LLM to subsequently generate writing assistance. The *Personalization* condition accumulates chunks in a different way, relying on the text passages selected by the user in real-time while reading. During the highlighting process, users can precisely manage their pool of raw materials. The original intention of this design is to reduce the interference of other non-essential information during content generation and to give users a sense of control through precise selection [51]. Subsequently, in the data processing phase (B), each chunk undergoes the same processing, being vectorized through an embedding model and stored in a vector database. The metadata of each chunk in the original text, including page location, unique identifiers, etc., is also stored to facilitate subsequent retrieval and relocation. When writers need AI assistance (C), their prompts are vectorized using the same embedding model, and candidate chunks are obtained through semantic similarity matching. In each retrieval, the default top 10 re-ranked chunks will be used for the suggestion generation[6].

*3.5.2 User Interface and Interaction Flow.* Fig. 2 shows the interface of the platform, consisting of four main components:

(1) **Reading Panel:** Participants begin by reviewing a preloaded source document. They can navigate through the document, highlight sections of text, or switch to another document. All highlights are stored in the real-time backend vector database and are used later for AI-generated suggestions.
(2) **AI Suggestion Panel:** After transitioning to the writing phase, users can select from one of the 2×6 available prompt templates (Table 1), customize the placeholder text, and receive suggestions from AI. Each generated suggestion contains one or several chunks. Each chunk includes one or several sentences.
(3) **Source Card List:** When clicking on a chunk, the corresponding sources are shown on the left panel. Clicking on a source card allows users to jump back to the original paragraph in the PDF viewer, facilitating a quick review.
(4) **Writing Panel:** Users draft their argumentative essay in a rich-text editor. Suggestions can be reviewed and inserted directly into the draft. Upon completion, users submit their work through the interface.

*3.5.3 AI-generated Writing Suggestion and Prompt Control.* During the argumentative writing stage, users can interact with our controlled prompting AI system to receive writing suggestions. The system supports six types of rhetorical elements based on the Toulmin model [43], which is a framework for analyzing and constructing arguments. It breaks down an argument into six key elements: claim, grounds, warrant, backing, qualifier, and rebuttal. For each element, we provided participants with two predefined prompt templates, shown in Table 1. Participants were instructed to modify only the placeholder text (e.g., "[claim]" or "[topic]") to ensure consistency across responses. Here, we provide an example of how to validly modify the prompt based on templates. In this example, the participant was attempting to construct a claim about *school uniforms*.

> Prompt template: *"What are 2-3 strong reasons from the reading that can support my claim: [claim]"*
> Example of a participant-modified prompt: *"What are 2-3 strong reasons from the reading that can support my claim: [wearing uniform in school reduced the crime rate]"*

Once the prompt has been submitted, the system will check whether the prompt is valid (see Table 3). When receiving an invalid template, the system will return a JSON response with an error message and a suggestion for the correct template, to remind the user how to use it correctly. Otherwise, the system will initiate the writing suggestion, retrieving the most relevant content via vector search and reranking. For the *Personalization* condition, only these highlights are used as the retrieval context in the LLM prompt, ensuring that the AI output is tightly linked to the participant's own reading behavior, whereas in the *Baseline* condition, the system would retrieve based on all reading materials. These retrieved candidates are subsequently reranked through Cohere's *rerank-v3.5* model, which identifies the top ten most relevant chunks for grounding the response. The system then formats these chunks, together with their metadata, such as unique identifiers and original locations, into the [CONTEXT_BLOCK] of Table 4, combines it with the validated user prompt, and forwards the complete instruction to GPT-4.1. The model generates a coherent paragraph that integrates only the provided evidence and appends [refs:...] tags that indicate which chunks contributed to each sentence. Finally, the system frontend parses the generated output and displays clickable source cards for each referenced chunk, allowing participants to return to the exact passages in the original articles and maintain transparency in how evidence supports the AI-generated suggestions.

---

[1] https://vuejs.org/
[2] https://www.djangoproject.com/
[3] https://us-east-1.console.aws.amazon.com/ec2/
[4] https://platform.openai.com/docs/models/gpt-4.1
[5] https://www.trychroma.com/

[6] The selection of the system parameters, such as chunk size and the number of retrieval chunks, is derived from multiple pilots and by referencing the implementation of other RAG systems.

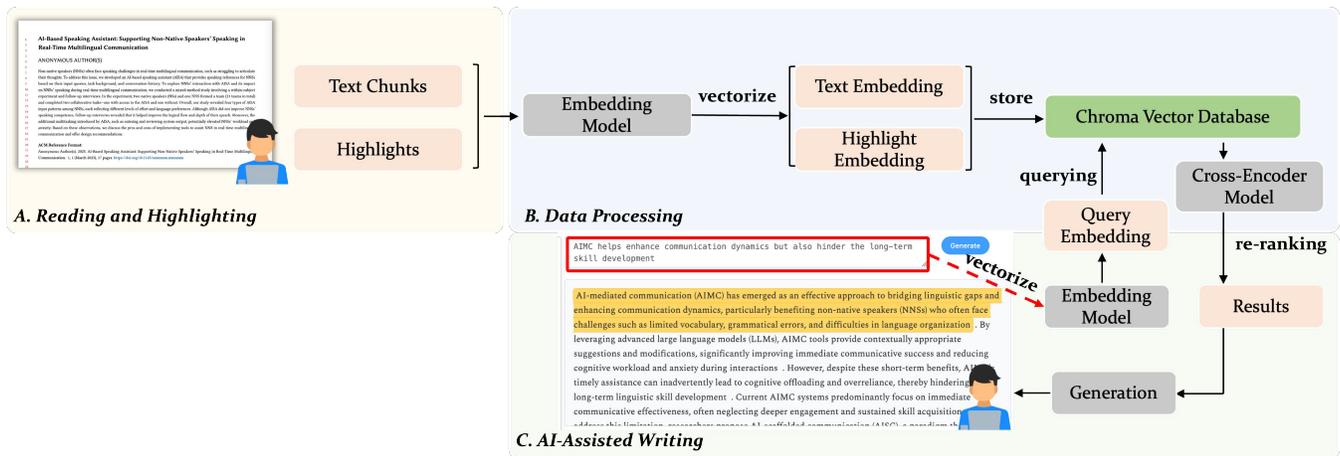

Figure 1: The technical architecture diagram linking reading behavior to personalized AI-assisted writing. User highlights (*Personalization* condition) and text chunks (*Baseline* condition) are separately vectorized and stored in a vector database according to the assigned conditions. During writing, queries are embedded to retrieve the most relevant segments, which are then re-ranked for performance improvement. These segments are finally used to generate writing suggestions.

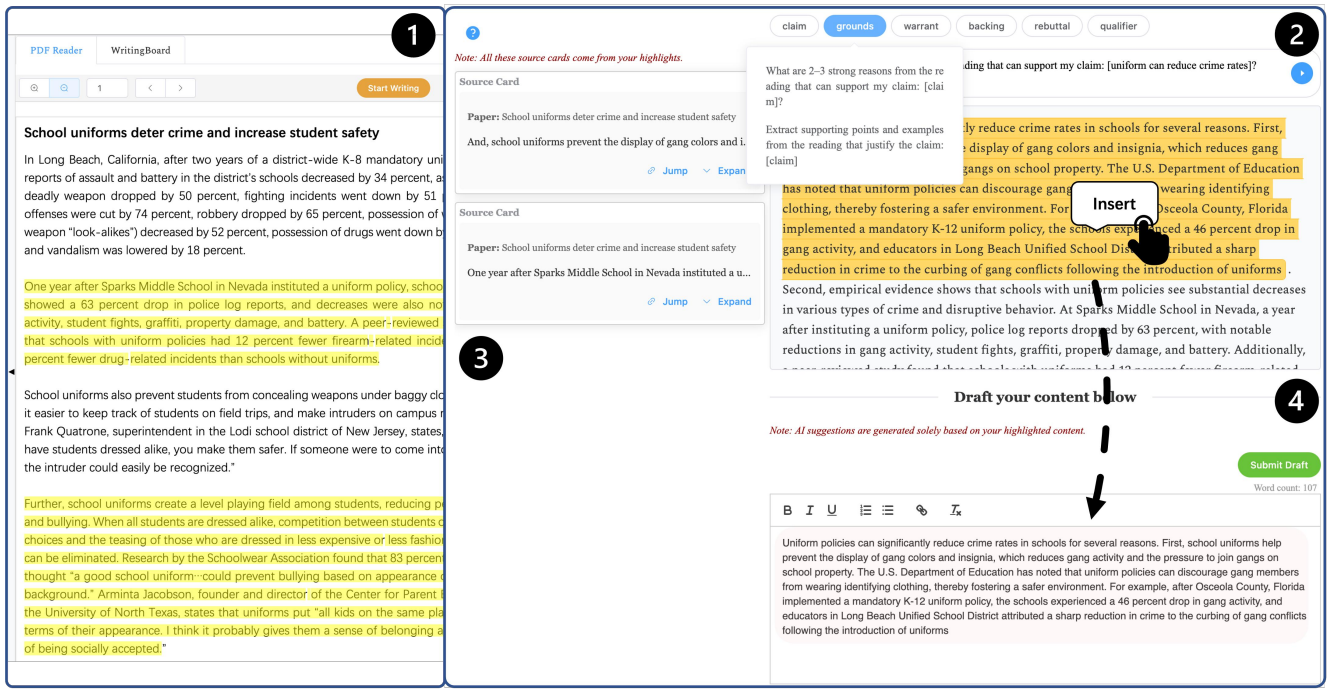

Figure 2: The interface includes four main components. (1) Reading panel allows reading and highlights. (2) The AI suggestion panel enables writers to compose prompts and get suggestions from AI. From top to bottom are the prompt construction area and the AI suggestion display area. Writers can choose from six preset prompt templates (12 templates in total) and modify them. Subsequently, AI-generated suggestions were displayed below the prompt area. (3) Source cards offer sources used in AI suggestions, which is convenient for writers for a quick review. (4) The writing panel is an area for writers to draft their argumentative essays.

## 3.6 Measurements

To address our research questions, we collected behavioral and self-report data through system logs and Likert scales. All measurement metrics are listed in Table 2. The symbol "#" denotes "the number of".

*3.6.1 Manipulation Check.* To verify that participants understood the mechanism of AI suggestion generation, we included a manipulation check item after the writing task: *"Please indicate your understanding of how the AI suggestions were generated in your writing task,"* with three options: (1) based on my highlights, (2) based on the full content, and (3) not sure. We excluded data from four participants because their responses was *"not sure"* or did not align with their assigned condition.

*3.6.2 Reading Process (RQ1, RQ3).* We monitored reading and highlighting behaviors through system logs. Subsequently, we derived the number of highlights, the density of highlights, and the average highlight length. We describe each measure as follows.

> ***number of highlights:*** how many times a participant highlights a part of text in the task.
> ***highlight density:*** #*highlighted words* / #*words of reading materials*.
> ***average highlight length:*** #*highlighted words* / *number of highlights*.

*3.6.3 AI-assisted Writing Process (RQ2, RQ3).* We monitored writing behaviors with AI through system logs and calculated the following indicators.

> ***number of prompts:*** how many times participants request AI suggestions.
> **#*generated AI chunks***: total number of chunks generated during AI-assisted writing.
> **#*accepted AI chunks***: number of AI-generated chunks accepted by the participant.
> **#*accepted AI words***: total number of words contained in the accepted chunks.
> **#*human-revised words***: number of words originally from AI that were revised by the participant.
> **#*AI-remaining words***: number of words originally from AI that remained unchanged in submission.

> ***acceptance ratio:*** #*accepted AI chunks* / #*generated AI chunks*.
> ***number of edits:*** times the participant made additions, deletions, and modifications made by the participant.
> ***revision intensity:*** #*human-revised words* / #*accepted AI words*.
> **#*submission words***: total number of words in submission.
> ***AI reliance:*** #*AI-remaining words* / #*submission words*.

*3.6.4 Perceived Mental Effort (RQ1, RQ3).* To comprehensively examine the subjective perceptions of the participants about mental effort throughout the reading and writing processes, we adapted previous scales on mental effort and engagement [15, 57, 58]. We asked participants to separately rate the following scales for the reading and writing stages on a 7-point Likert scale (1 = strongly disagree, 7 = strongly agree). For each scale, the average score served as the measure of perceived effort. The items of each scale are as follows.

**Reading Effort.** *"I feel mentally demanding when identifying the critical information in the reading articles." "Overall, I feel mentally demanding when completing the reading."* Cronbach's alpha $\alpha = 0.920$ indicated the reading effort scales had excellent internal consistency.

**Writing Effort.** *"I feel mentally demanding when proposing ideas and claims" "I feel mentally demanding when organizing and integrating ideas." "I feel mentally demanding when editing and revising the text during the writing task." "Overall, I feel mentally demanding when completing the writing."* Cronbach's alpha $\alpha = 0.939$ indicated the writing effort scales had excellent internal consistency.

*3.6.5 Perceptions in AI-Assisted Writing (RQ2, RQ3).*

*Sense of Ownership.* We adapted prior work on writing ownership [4, 41, 93] and asked the following items: *"I feel like I am the author of the text." "I am the main contributor to the content of the resulting text." "I am totally accountable for the text."* Responses were recorded using a 7-point Likert scale (1 = strongly disagree; 7 = strongly agree). The averaged score served as the ownership perception score. Cronbach's alpha $\alpha = 0.907$ indicated the ownership scales had excellent internal consistency.

*Sense of Autonomy.* We adapted autonomy scales in [20, 62] and asked the following items: *"I felt like I was writing the text and the AI was assisting me." "The writing outcome aligned with my intention and plan." "I had full control over the writing content."* Participants rated these items on a 7-point Likert scale (1 = strongly disagree; 7 = strongly agree); the averaged score served as the autonomy perception score. Cronbach's alpha $\alpha = 0.819$ indicated the autonomy scales had good internal consistency.

*Credit Attribution.* We measured perceived contributions from both the participant and the AI [62]. Specifically, we asked participants to rate two sets of question items on a 7-point Likert scale (1 = strongly disagree; 7 = strongly agree) about the extent of their / AI's contributions to idea generation, organization, phrasing, and overall content. The items are:

> **Self-Credit Attribution.** *"I feel I contributed to the core ideas in the writing." "I feel I contributed to the organization and structure of the writing." "I feel I contributed to the wording* and phrasing of the writing." "Overall, I contribute to the writing." We found that the self-credit scales had good internal consistency (Cronbach's $\alpha$ = 0.849).

Table 1: Prompt templates used in the AI Suggestion Panel, categorized by rhetorical element [43]. Each prompt contains a placeholder (i.e., [topic] or [claim]) that participants were instructed to fill in.

| Type | Prompt Template |
| --- | --- |
| Claim | - Based on the reading, what is a clear and arguable claim I can make about the topic: [topic]?<br>- Generate a concise and debatable thesis statement on topic: [topic], using evidence from the reading. |
| Grounds | - What are 2–3 strong reasons from the reading that can support my claim: [claim]?<br>- Extract supporting points and examples from the reading that justify the claim: [claim]. |
| Warrant | - Explain how the evidence from the reading supports the claim: [claim]. What underlying assumption links them?<br>- Make the reasoning explicit: why do the points from the reading justify this claim: [claim]? |
| Backing | - Find additional background, theories, or expert opinions in the reading that strengthen the reasoning for the claim: [claim].<br>- What contextual or scholarly evidence in the reading can further reinforce the logic of the claim: [claim]? |
| Rebuttal | - What possible counterarguments can be raised against the claim: [claim], based on the reading, and how can I respond to them?<br>- Generate a paragraph that acknowledges and rebuts opposing the claim: [claim], using reading evidence. |
| Qualifier | - Help me refine my claim: [claim] with qualifiers like "in most cases" based on nuances in the reading.<br>- According to the reading, under what conditions does my claim: [claim] hold true? Reformulate it with appropriate qualifiers. |

**AI-Credit Attribution.** "I feel AI contributed to the core ideas in the writing." "I feel AI contributed to the organization and structure of the writing." "I feel AI contributed to the wording and phrasing of the writing." "Overall, I feel AI contributed to the writing." Cronbach's alpha $\alpha$ = 0.755 suggested the AI-credit scales had acceptable internal consistency.

Each set of items (Self-credit and AI-credit) was averaged to represent the participant's perceived credit attribution to self and AI.

*Satisfaction with Writing.* We referred to [18, 78] and developed a scale to evaluate participants' satisfaction. We asked the participants to rate the following items on a 7-point Likert scale (1 = Very Low; 7 = Very High): *"How much did you enjoy this writing task?" "How satisfied are you with the writing outcome?" "How satisfied were you with the AI assistance?"* We found the satisfaction scales had excellent internal consistency (Cronbach's $\alpha$ = 0.913).

*Alignment.* To the best of our knowledge, very limited existing validated scales capture participants' perceptions of alignment between AI-generated outputs and their own ideas or expectations in the writing task context. Therefore, we developed a simple 7-point scale (1 = Strongly Disagree; 7 = Strongly Agree) to assess how writers perceive the alignment of AI-generated content with their own thoughts and expectations. *"I feel the AI outputs align with my expectations of what AI should generate." "I feel the AI outputs align with my ideas." "Overall, I believe AI understands my intentions well."* All items were averaged as the final indicator. Cronbach's alpha $\alpha$ = 0.939 indicated excellent internal consistency of the alignment scales.

*AI Performance.* Similarly, given the lack of existing measures tailored to AI performance under our experiment settings, we devised a task-specific scale to evaluate participants' perceived AI performance. We developed a 7-point scale (1 = Strongly Disagree; 7 = Strongly Agree) involving items: *"The AI system was capable." "The AI system provided quality responses." "The AI system provided responses aligned with my expectations." "The AI system was trustworthy and reliable in helping me complete the task." "The AI system had a positive impact on my completion of the task."* The mean of all items was used as the final indicator. We found the AI performance scales had excellent internal consistency (Cronbach's $\alpha$ = 0.935).

## 4 Results

This section presents the impacts of using reading highlights for personalized AI-assisted writing on reading behaviors and perceptions (RQ1), writing behaviors and perceptions (RQ2). In addition, it also reveals the relationship between reading and AI-assisted writing (RQ3). In the following section, **Personalization** refers to the condition where participants' highlights are integrated when AI generates writing suggestions, while **Baseline** refers to the condition where AI generates without personalization, solely based on the reading materials.

Table 2: Summary of measurements used in the study. We denote $\alpha$ as Cronbach's alpha for the scale reliability.

| Category | Measure | Description / Formula | Reliability |
| --- | --- | --- | --- |
| Reading Behaviors | Number of highlights | Count of highlight actions in reading | System log |
| | Highlight density | #highlighted words / #total words | System log |
| | Average highlight length | #highlighted words / #highlights | System log |
| Writing Behaviors | Number of prompts | Times participants request AI suggestions | System log |
| | Generated AI chunks | Total chunks generated | System log |
| | Accepted AI chunks | Number of AI chunks inserted into text | System log |
| | Accepted AI words | Words contained in accepted chunks | System log |
| | Human-revised words | AI words revised by participant | System log |
| | AI-remaining words | AI words retained without revision | System log |
| | Acceptance ratio | #accepted AI chunks / #generated AI chunks | Derived metric |
| | Number of edits | Total additions, deletions, modifications | System log |
| | Revision intensity | #human-revised words / #accepted AI words | Derived metric |
| | Submission words | Total word count of final essay | System log |
| | AI reliance | #AI-remaining words / #submission words | Derived metric |
| Mental Effort | Reading effort | 2 items (7-pt Likert) | $\alpha = 0.92$ |
| | Writing effort | 4 items (7-pt Likert) | $\alpha = 0.94$ |
| Subjective Perceptions | Sense of ownership | 3 items (7-pt Likert) [41, 93] | $\alpha = 0.91$ |
| | Sense of autonomy | 3 items (7-pt Likert) [20, 62] | $\alpha = 0.82$ |
| | Self-credit attribution | 4 items (7-pt Likert) [62] | $\alpha = 0.85$ |
| | AI-credit attribution | 4 items (7-pt Likert) [62] | $\alpha = 0.76$ |
| | Satisfaction with writing | 3 items (7-pt Likert) [18] | $\alpha = 0.91$ |
| | Alignment | 3 items (7-pt Likert) | $\alpha = 0.94$ |
| | AI performance | 5 items (7-pt Likert) | $\alpha = 0.94$ |

We used JASP v0.18.3[7] to analyze the data. For all data based on Likert scale responses, we used non-parametric methods (Mann-Whitney U test) for analysis, as these responses are ordinal in nature and do not necessarily meet the assumptions of parametric tests. For other log data, we employed Welch's t-test for between-group comparisons. Unlike the classical Student's t-test, Welch's test does not assume equal variances between groups and is more robust when sample sizes or variances are unequal. Finally, we conducted a correlation analysis for each condition, aiming to determine the impact of the reading phase on AI-assisted writing under various conditions.

## 4.1 The Effect of Personalization on Reading Process (RQ1)

*4.1.1 Engagement in Reading.* The results showed that using highlights for personalization encouraged writers to spend more time engaging in reading, where they significantly highlighted more and at a higher density, but they did not perceive it as more effortful.

The Welch's t-test (see Fig. 3) showed that using highlights for personalized AI-assisted writing caused a significant effect on highlighting behaviors. Specifically, participants in the *Personalization* condition made significantly **more highlights** ($t(41.470) = -4.610, p < 0.001, d = -1.351$), had a higher **highlight density** ($t(27.234) = -9.081, p < 0.001, d = -2.631$), and had longer **average highlight length** ($t(26.418) = -3.058, p = 0.005, d = -0.885$). Participants in the *Personalization* highlighted significantly more times ($M = 22.958; SD = 13.437$) than those in *Baseline* ($M = 7.227; SD = 9.522$), with greater highlight density ($M = 0.452; SD = 0.207$ vs. $M = 0.051; SD = 0.060$) and longer highlight length ($M = 60.051; SD = 74.156$ vs. $M = 12.062; SD = 19.447$). In terms of **reading duration**, participants in *Personalization* also spent significantly more time on reading ($t(42.184) = -2.491, p = 0.017, d = -0.731$), with an average of $M = 569.741; (SD = 275.546)$ seconds compared to *Baseline* ($M = 392.718; SD = 203.759$ seconds).

The Mann-Whitney U test showed that there was no significant difference ($U = 297.500, p = 0.462, r = 0.127$) in the perceived reading effort between *Personalization* ($M = 4.125; SD = 1.245$) and *Baseline* ($M = 4.568, SD = 1.391$).

---
[7]https://jasp-stats.org/

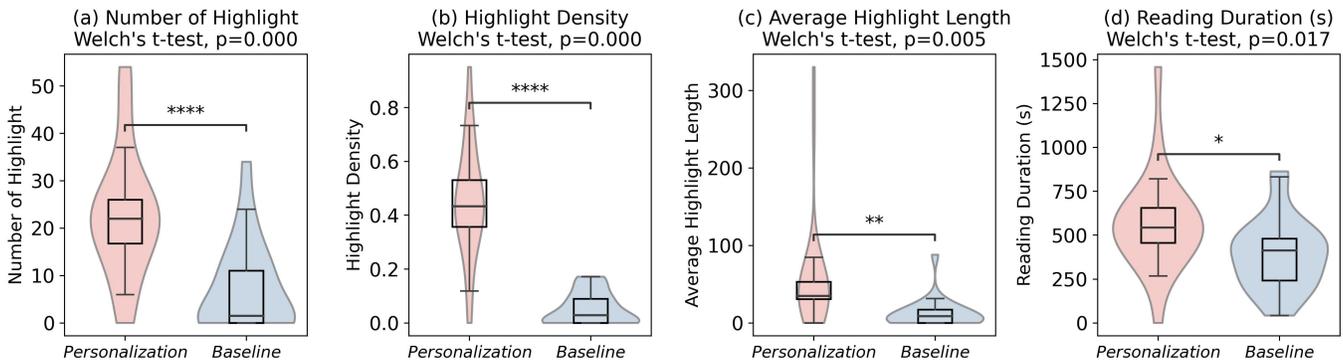

Figure 3: Comparison between *Personalization* and *Baseline* conditions in reading behaviors: (a) Highlight Count, (b) Highlight Density, (c) Average Highlight Length, (d) Reading Duration. The violin plots represent the distribution, while the overlaid boxplots show the median and interquartile ranges. Significant differences between the conditions are indicated by p-value annotations, where * < 0.05, ** < 0.01, *** < 0.001, **** < 0.0001.

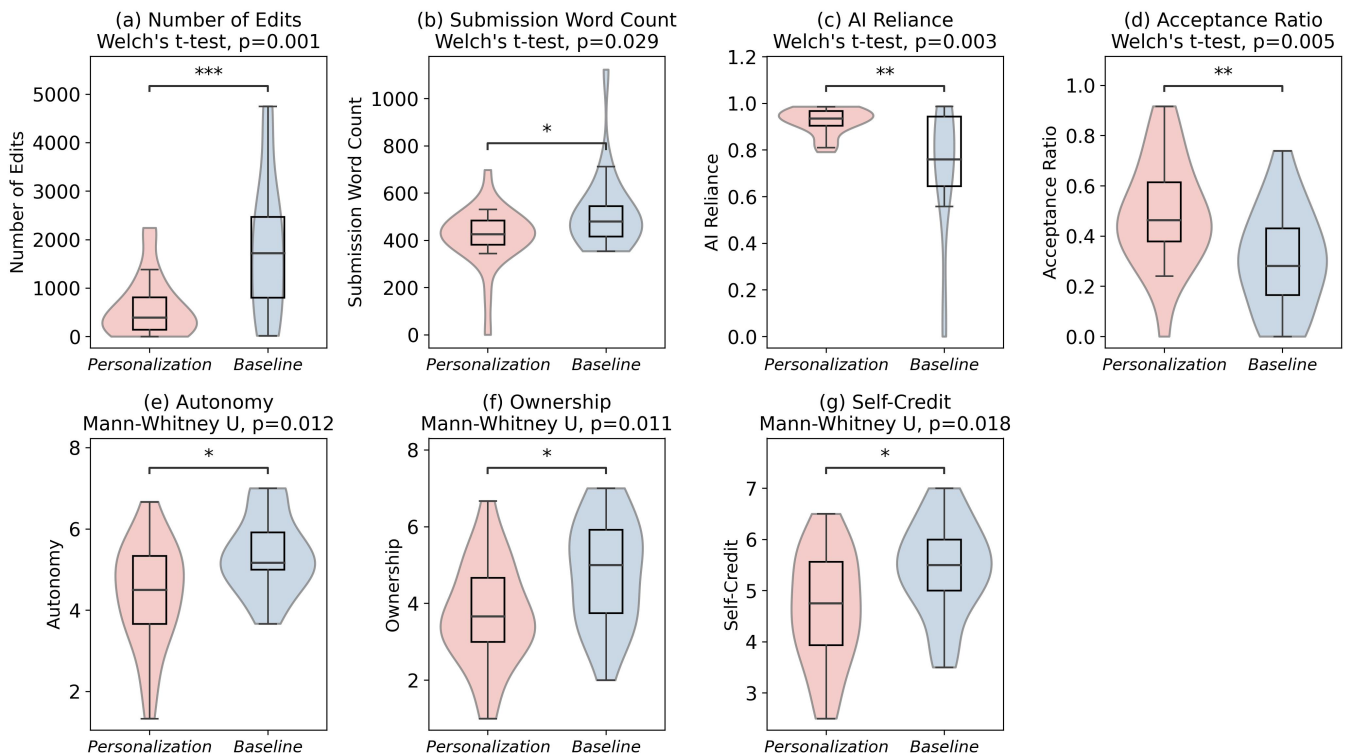

Figure 4: Comparison between *Personalization* and *Baseline* in writing behaviors and perceptions: (a) Number of Edits, (b) Submission Word Count, (c) AI Reliance, (d) Acceptance Ratio, (e) Autonomy, (f) Ownership, (g) Self-Credit. The violin plots represent the distribution, while the overlaid boxplots show the median and interquartile ranges. Significant differences between the conditions are indicated by p-value annotations, where * < 0.05, ** < 0.01, *** < 0.001.

## 4.2 Writing Behaviors and Perceptions (RQ2)

Our results indicated that when using highlights for personalized AI-assisted writing (*Personalization*), participants showed greater reliance on AI and higher acceptance of AI output, but also engaged less actively in editing and produced shorter drafts. Importantly, their subjective perceptions of autonomy, ownership, and self-credit are significantly lower compared to those in the *Baseline* group.

*4.2.1 Engagement in Writing.* The Welch's t-test revealed several significant differences between the two conditions (see Fig. 4 (a,b)). First, participants in the *Baseline* group had a significantly higher ***number of edits*** ($t(28.040) = 3.753, p < 0.001, d = 1.124$) compared with those in the *Personalization* group. On average, the *Baseline* group made $M = 1847.136$ edits ($SD = 1454.245$), whereas the *Personalization* group made only $M = 587.958$ edits ($SD = 627.752$). Secondly, the ***submission word count*** was significantly higher ($t(39.500) = 2.265, p = 0.029, d = 0.672$) in the *Baseline* group ($M = 514.636; SD = 165.454$) compared to the *Personalization* group ($M = 415.208; SD = 127.962$). We did not detect a significant difference in terms of writing duration ($t(41.868) = 1.613, p = 0.114, d = 0.478$).

*4.2.2 Reliance on AI.* The Welch's t-test (see Fig. 4 (c, d))revealed that participants in *Personalization* exhibited significantly more ***AI reliance*** than those in *Baseline* ($t(22.460) = -3.339, p = 0.003, d = -1.006$). The average AI reliance score in *Personalization* was $M = 0.924; SD = 0.056$, compared to $M = 0.710; SD = 0.295$ in *Baseline*. Similarly, the ***acceptance ratio*** was significantly higher ($t(43.939) = -2.952, p = 0.005, d = -0.870$) in *Personalization* ($M = 0.489; SD = 0.214$) than in *Baseline* ($M = 0.307; SD = 0.203$). The difference in the accepted chunk count was not significant ($t(36.661) = -1.835, p = 0.075, d = -0.546$), but we observed a trend of accepting more chunks in *Personalization* ($M = 11.083; SD = 4.303$) than in *Baseline* are $M = 8.136; SD = 6.304$.

No significant group differences were observed for the number of prompts ($t(43.980) = 1.222, p = 0.228, d = 0.360$), revision intensity ($t(21.029) = 1.288, p = 0.212, d = 0.388$).

*4.2.3 Sense of Autonomy, Ownership, and Credit Attribution.* A Mann–Whitney U test (see Fig. 4 (e, f, g))indicated significant group differences in ***autonomy*** ($U = 378.500, p = 0.012, r = 0.434$), ***ownership*** ($U = 379.000, p = 0.011, r = 0.436$), and ***self-credit*** ($U = 371.500, p = 0.018, r = 0.407$). Participants in *Baseline* reported higher autonomy ($M = 5.379; SD = 0.950$) compared to those in *Personalization* ($M = 4.389; SD = 1.273$). Similarly, ownership was higher in *Baseline* ($M = 4.803; SD = 1.402$) relative to *Personalization* ($M = 3.722; SD = 1.321$). Likewise, participants in the *Baseline* group reported higher self-credit scores ($M = 5.420; SD = 0.927$) than those in *Personalization* ($M = 4.635; SD = 1.093$).

No significant group differences were observed for writing effort ($U = 282.000, p = 0.699, r = 0.068$), AI performance ($U = 165.000, p = 0.284, r = 0.222$), AI credit ($U = 266.500, p = 0.965, r = 0.009$), satisfaction ($U = 317.500, p = 0.240, r = 0.203$), and alignment ($U = 239.500, p = 0.155, r = 0.267$).

## 4.3 The Relationship Between the Reading and Writing Processes (RQ3)

The above results (RQ1 and RQ2) show significant differences in reading and writing metrics across different conditions: participants in *Personalization* did more and denser highlights with fewer editing activities, while also exhibiting lower autonomy, ownership, and self-credit. This may indicate that the relationships between reading and writing in the two conditions are different. In the AI-assisted writing context, there was limited prior empirical evidence or hypotheses focusing on reading's influence. Therefore, we employed correlation analysis to broadly explore the relationships between measures in the reading phase and those in the writing phase.

Specifically, in the reading stage, we focus on the perceived *reading effort*. Between the two indicators of highlighting, *number of highlight* and *highlight density*, we ultimately selected *highlight density*, as it integrates both the number and the length of highlights, thus serving as a more representative measure of highlighting behavior.

Regarding the writing stage, variables reflecting writing behaviors involve *number of edits* [18], *revision intensity*, *accepted chunk count*, and *acceptance ratio*. In terms of writing perceptions, we analyzed *autonomy, ownership, self-credit,* and *AI-credit* [22, 83, 87], which are commonly used in prior literature to reflect the AI-assisted writing experience.

In addition, since previous research [59, 86] has shown that the alignment of AI content with human thoughts affects user satisfaction ratings of AI, we want to explore whether reading highlights influence the human-AI interaction experience. Therefore, we included writers' perceived *alignment, AI performance,* and *satisfaction with writing*.

*4.3.1 Summary of Findings.* The results indicate that when highlights were not used for personalization, reading effort and highlighting behaviors were positively reinforced, leading to increased writing engagement and more positive perceptions. In contrast, when highlights were integrated into personalization, these positive relationships weakened or reversed, suggesting that personalization altered the natural link between reading and writing, undermining autonomy, ownership, and satisfaction.

*4.3.2 Impact of Reading Effort.* In the *Baseline* group, correlation analyses (see Fig. 5a) revealed that when reading was independent of AI-assisted writing, participants' invested effort during reading directly translated into both higher cognitive input in writing and more positive perceptions of authorship and collaboration with AI.

Specifically, ***reading effort*** (RE) was significantly associated with multiple writing-related outcomes. Participants who invested greater RE subsequently reported higher ***writing effort*** (WE, $r = 0.782, p < 0.001$), greater ***autonomy*** (AU, $r = 0.502, p = 0.017$), and higher ***AI credit*** (AC, $r = 0.475, p = 0.026$). RE was also positively correlated with stronger ***alignment*** (AL, $r = 0.580, p = 0.012$).

In the *Personalization* condition, correlation analyses (see Fig. 5b) revealed that RE was again positively associated with ***writing effort*** (WE, $r = 0.643, p < 0.001$). However, RE did not exhibit significant associations with perceptions such as autonomy (AU), ownership (OW), or self-credit (SC).

*4.3.3 Impact of Highlighting Behaviors.* The findings indicated that positive highlighting behavior during reading was linked to more active writing engagement only when highlights were not used for personalization; otherwise, highlighting was associated with diminished writing perceptions.

In the *Baseline* group (see Fig. 5a), HD showed significant positive associations with ***number of edits*** (NE, $r = 0.465, p = 0.029$) and ***revision intensity*** (RI, $r = 0.447, p = 0.037$), meaning that participants in *Baseline* whose more active highlighting behaviors were closely associated with greater engagement in editing and revision.

By contrast, in the *Personalization* condition (see Fig. 5b), the role of highlighting shifted. Specifically, HD was positively correlated

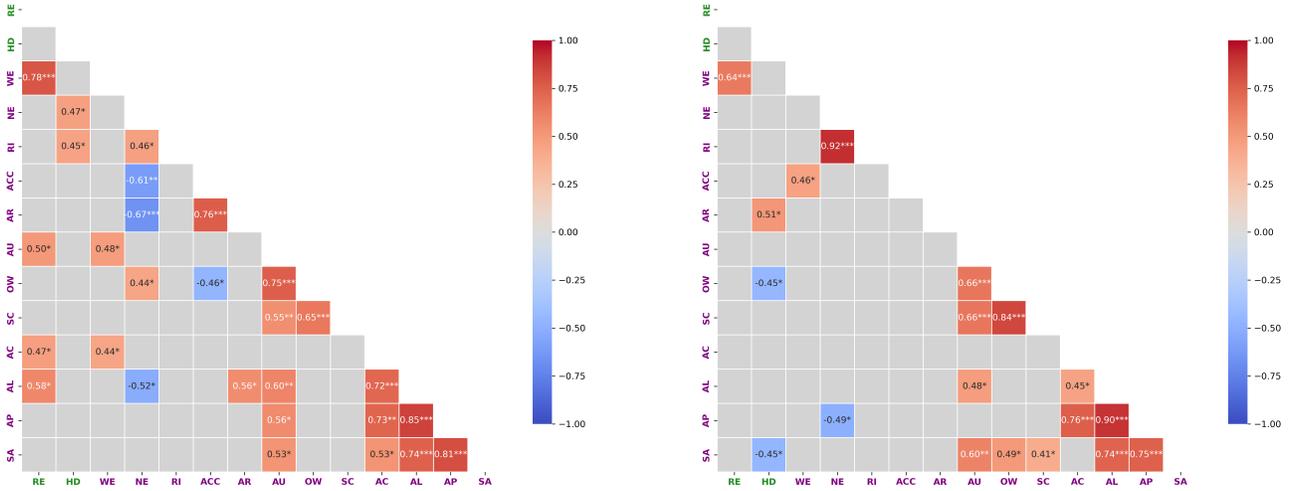

(a) Baseline condition. Reading effort (RE) and highlight density (HD) were positively correlated with writing behaviors and perceptions, including pairs of (RE, WE), (RE, AU), (RE, AC), (RE, AL), (HD, NE), and (HD, RI).

(b) Personalization condition. Compared to the baseline condition, correlations between reading and writing measures are reduced or shifted. RE is only positively correlated with WE (RE, WE), while HD is positively correlated with AR (HD, AR) and negatively correlated with writing behaviors and perceptions, including pairs (HD, OW) and (HD, SA).

Figure 5: Correlation matrices of reading and writing variables in *Baseline* and *Personalization* conditions. Significant correlation effect sizes are shown in red–blue (indicating direction and magnitude), with significance levels: $^{*}p < 0.05$, $^{**}p < 0.01$, $^{***}p < 0.001$. For simplicity, non-significant correlations are shown in grey. For readability, measure abbreviations are also color-coded: RE = Reading Effort; HD = Highlight Density (Reading measures), and WE = Writing Effort; NE = Number of Edits; RI = Revision Intensity; ACC = Accepted Chunk Count; AR = Acceptance Ratio; AU = Autonomy; OW = Ownership; SC = Self-Credit; AC = AI Credit; AL = Alignment; AP = AI Performance; SA = Satisfaction (Writing measures).

with *acceptance ratio* (AR, $r = 0.512, p = 0.011$), but negatively correlated with *ownership* (OW, $r = -0.448, p = 0.028$) and *satisfaction* (SA, $r = -0.455, p = 0.026$).

## 5 Discussion
### 5.1 Overview

Our research revealed how highlights can serve as personalization cues for AI-assisted writing. Specifically, we found that personalization facilitated more engagement during the reading phase **(RQ1)** (Section 4.1.1). However, it negatively influenced the writing phase **(RQ2)**, leading to higher reliance on AI (Section 4.2.2), reduced engagement in writing (Section 4.2.1), and diminished perceptions of autonomy, ownership, and credit attribution (Section 4.2.3). Further analysis revealed that the relationship between reading and writing also differed across conditions **(RQ3)**: in the *Personalization* condition, active engagement in reading did not translate into benefits for writing; in contrast, in the *Baseline* group, reading behaviors and perceptions were positively correlated with writing behaviors and perceptions (Sections 4.3.2 and 4.3.3). In the following, we discuss these findings and derive design implications for more effective personalization in AI-assisted writing.

### 5.2 Personalized AI Increases Superficial Engagement, Reliance, and Diminishes Writing Perceptions

Our results indicate that although personalization promotes more highlighting, users do so not to enhance their understanding of the reading text, but rather to provide input for AI to obtain better and richer AI-generated suggestions. Quantitatively, participants in the *Personalization* condition made more highlights and spent more time in the reading stage. The highlight density of the *Personalization* group (M = 0.452; SD = 0.207) was about 10 times that of the *Baseline* group (M = 0.051; SD = 0.060). However, this highlighting is more like an adaptive strategy, which has been widely found in other scenarios such as users' behavior adjustment in recommendation systems for better results [12, 13, 30]. In this superficial engagement, highlighting may lose its original function in deepening understanding and distinguishing texts with various importance [52, 96]. Instead, it becomes a means to ensure AI obtains sufficient context for suggestion generation.

This personalization process aligns with the notion of *extrinsic personalization* [6], derived from education research, which explains a phenomenon where learners engage strategically with a system for external rewards or benefits, rather than for authentic learning. Correspondingly, "feeding AI for easy-to-use writing suggestions" acts as an external motivation for personalization behavior, turning the original personalization process into a "mechanical process" of repeatedly adding highlights.

In addition to changing writers' reading behavior, our results showed that personalization also increased participants' reliance on AI-generated content during writing. Writers in the *Personalization* group exhibited a significantly higher acceptance rate of AI suggestions (Section 4.2.2). In many cases, they made only minor modifications, with the total number of edits being roughly one-third of that in the *Baseline* group. Their final submissions were also shorter, averaging about 100 fewer words. This aligns with Hwang et al. [34], who noted that while many authors prefer personalized AI writing tools, they worry that such personalization could make them more likely to accept AI suggestions. This, in turn, may allow AI to exert a greater influence on their writing. In a similar vein, our findings provide empirical evidence of the risk of overreliance due to personalization in AI-assisted writing. Correspondingly, *Personalization* participants reported less autonomy, ownership, and self-credit.

Ontanon and Zhu [56] introduced the concept of the "personalization paradox," which suggests that while personalization is designed to tailor digital experiences to align with individual interests and preferences, it serves to influence and direct one's daily decisions and behaviors, diminishing the range of visible options and consequently limiting personal autonomy. We have extended the application of this paradox in the context of AI-assisted writing, where a personalized design can influence the motivation and engagement in reading, while undermining user autonomy and ownership in writing.

Taken together, our findings suggest that a seemingly effective personalized design may actually have an unexpectedly negative impact. The possible reasons for this are that it triggers different user motivations, emphasizing the importance of personalized designs in guiding behavioral motivation.

### 5.3 Personalization Changes the Functions of Reading in AI-Assisted Writing

Reading and writing are interconnected cognitive processes [45, 79]. The correlation analyses of our measures in the *Baseline* condition (Fig. 5a) empirically demonstrated the relationship between reading and writing in the context of AI-assisted writing. In the *Baseline* condition, the reading effort was positively correlated with the writing effort. Moreover, active highlighting behaviors correlated with the active editing behaviors in writing. These findings, therefore, offer empirical evidence that future AI-assisted writing systems can promote an engaged writing experience by focusing on the reading process. Moreover, the reading effort was also correlated with participants' autonomy, which is a core perception in writing [25, 76]. This suggests that future designs can help maintain user autonomy in writing tasks through the design of the reading phase, expanding existing research on ways to maintain autonomy [54, 62].

However, the correlation analyses for the *Personalization* condition revealed a different relationship. We found a significant negative correlation between highlighting behavior and ownership, while a positive correlation with acceptance ratio (see Fig. 5b). This indicates that personalization disrupts the contribution of reading to writing, which was originally found when writers engaged with AI without personalization (in *Baseline*). One possible explanation is related to the fact that extrinsic personalization alters behavioral motivation [6, 66]. In the *Baseline* condition, reading and highlighting are to deepen their understanding and distinguish the important content [52], therefore, the more reading effort writers invest, the more they would feel a sense of control over the final writing results, and would be more willing to edit the writing content. In contrast, the highlighting motivation of *Personalization* participants is rather driven by extrinsic personalization, i.e., external factors of providing information to AI. Under this goal, highlighting aims to create richer content for personalized writing suggestions. This naturally leads to a higher acceptance ratio and corresponds to lower ownership. In summary, personalization seems to change the function of reading in the AI-assisted writing process from comprehending the content to "feeding the AI."

### 5.4 Design Implications

Overall, our findings reveal that personalization reshapes reading and writing in AI-assisted writing, which also suggests a change in users' underlying motivations, often undermining autonomy and ownership. These insights highlight the need to carefully design personalization in AI-assisted writing, which we address through the following design implications.

*5.4.1 Cultivating Intrinsic Motivation through Personalization.* Our study shows that personalization may unintentionally shift writers' motivation from intrinsic comprehension to extrinsic goals such as "feeding the AI with enough inputs." To counter this, personalization should focus on strengthening intrinsic goals. For instance, systems can embed reflective reading [5, 85], by integrating note-taking [9, 40], mind-mapping tools [29], or reflective prompts[8], or require evidence-mapping before AI suggestions appear before AI suggestions appear, positioning the AI as a cognitive partner rather than a shortcut. These scaffolds help maintain autonomy [25, 76], reinforcing the writer's role in directing the process and preserving authentic ownership [25, 51, 55, 83].

*5.4.2 Preserving the Cognitive Role of Highlights.* Highlights, as externalizations of cognition [64, 71], can be a powerful signal for personalization. However, in our study, participants in the *Personalization* condition often over-highlighted, which diluted the ability to distinguish between important and unimportant text. To address this, systems should guide users toward selective and meaningful highlighting. For example, future design can explore automated feedback in the reading process [75], which can flag when highlights are too broad, as well as use post-reading self-explanations [47] to ensure highlights serve comprehension. Importantly, systems could also separate "reading highlights" (for comprehension) from "personalization cues" (for AI input), and explicitly nominate a subset as *personalization cues*. This preserves the cognitive role of highlights as externalized cognition [64, 71] while ensuring that only deliberate data steer the model. Such operations can also encourage the personalization process to incorporate more deliberate choices by the user [17], thereby enhancing the development of autonomy and ownership [61, 69].

---

[8]https://lib.d.umn.edu/research-collections/find-what-read-next/reflection-questions

## 6 Limitations and Future Work

We acknowledge several limitations of our study. First, we started with argumentative writing as the typical source-based writing task for studying personalized AI-assisted writing as a first step. We acknowledge that writing genres vary substantially in their cognitive structures. Forms such as creative, reflective, or narrative writing, which do not require evidence synthesis or source consultation, may elicit different personalization effects. Future research should therefore examine how highlight-based personalization operates across a broader set of genres.

Second, our personalization feature was limited to highlights; other forms of reading annotations, such as comments, are still underexplored. We encourage future research to explore other aspects of reading interaction as a personalization cue.

Third, participants in the current experiment setup were explicitly informed that their highlights determined the personalized suggestions. Other awareness settings, such as implicit or partially transparent personalization, may lead to different engagement patterns. Future research can systematically vary awareness levels to understand how transparency interacts with personalization behavior and its outcome.

Additionally, the curated topics and readings may also shape how participants engage with the task. Although we selected comparable New York Times articles and instructed participants to rely solely on the provided texts, differences in topic interest, familiarity, and preexisting beliefs may still influence satisfaction, highlighting behaviors and writing strategies. Self-reported interest may also be affected by self-presentation [80] or preference falsification [46], and prior beliefs can introduce confirmation bias during reading [10, 35]. Future work should more systematically account for these topic-related individual differences when examining personalized AI-assisted writing.

We also believe that participants' familiarity with argumentative writing can influence how they read, write, and engage with the AI assistance. While we ensured that all participants had experience in argumentative writing, some participants may be more familiar with writing (e.g., they write as a profession). We encourage future work to investigate the role of task familiarity or expertise in AI-assisted writing.

## 7 Conclusion

This study examined how using reading highlights to personalize AI-assisted writing reshapes both reading and writing practices. While highlight-based personalization successfully promoted more active reading engagement, it paradoxically turned highlighting into an instrumental act of "feeding the AI," leading to greater reliance on AI output and diminished perceptions of autonomy, ownership, and self-credit. In contrast, when AI suggestions were generated from the full reading materials, reading effort and highlighting behaviors supported deeper engagement in writing and strengthened writers' sense of authorship. Our findings address the gap of existing studies by investigating how reading can be used for personalized AI-assisted writing, revealing its impact on writing behavior and perception while extending prior research by identifying an AI personalization paradox: designs that seem to enhance support may, in practice, undermine autonomy and ownership. We call for future AI writing systems to prioritize intrinsic personalization, actively explore, and utilize the positive role reading plays in personalization.

## Acknowledgments

This research was supported by the NUS Artificial Intelligence Institute (NAII), under Grant A-8003879.

## A Prompts

**Table 3: Prompt Validator Template.** Validates user prompts against allowed templates. [GROUNDTRUTH_PROMPTS] is replaced with permitted templates (see Table 1). [USER_PROMPT] is the user's actual input.

---

**Prompt Validator System Prompt**

You are a strict prompt validator for an AI-assisted writing tool.
Your task is to check whether a user's input prompt exactly follows one of the 12 allowed templates, with the only change being replacing the placeholder content. Placeholder text is defined using brackets, such as [topic] or [claim].
Here is the list of allowed templates: [GROUNDTRUTH_PROMPTS]
**Validation steps:**
(1) Identify if the user's prompt matches any template by replacing placeholders with actual content.
(2) If placeholder is completely unchanged (still keep [claim] or [topic]) return an error. Note that: **Keeping the square brackets does not necessarily mean that the placeholders have not been replaced, as long as the content in bracket has expressed user's own ideas**
(3) If the prompt doesn't match any pre-defined template, return an error.

**If valid, return this JSON:**
```
{
"type": "legal_prompt",
"matched_template": "<matched prompt template>",
"standardized_prompt": "<user's prompt>"
}
```
**If invalid, return this JSON:**
```
{
"type": "illegal_prompt",
"error": "<give detailed reason>",
"hint": "<Suggest closest template>"
}
```
**Now check this <user's prompt>:**
[USER_PROMPT]

---

## B Writing Evaluation

The goal of the study was to understand how personalization influences people's writing behavior regardless of the writing quality. To meet this goal, we adopted a relatively strict prompt design (see Section 3.5.3) when providing writing suggestions. However, this setting may prevent participants from fully using LLM in their own way for writing assistance, which could potentially affect the participants' writing quality. Nevertheless, we still provided an evaluation of the writing results as a supplementary side result. We recruited external raters who did not know the purpose of the study to assess participants' writing. The quadratic weighted kappa (QWK) is 0.532, which is an acceptable consistency [1]. The rubrics were adapted from [1] and finally involved 9 dimensions: Thesis, Claims, Evidence for Claims, Reasoning, Reordering/Organization, Rebuttal, Precision, Fluency, Conventions / Grammar / Spelling. Each item was scored on a scale of 1-4: "1-poor", "2-developing", "3-proficient", or "4-excellent". The essay score ranges from 9 to 36. The average of the two researchers' scores was used for data analysis.

**Table 4: Writing Assistant Template.** Generates paragraphs from document sources. [CONTEXT_BLOCK] contains source chunks; [USER_PROMPT] is the validated prompt. Each chunk_id refers to a unique identifier assigned to highlights or split raw content during PDF pre-processing. These identifiers allow the system to precisely locate and trace the exact portions of highlights or original content that contribute to the generated paragraph.

---

**Writing Assistant System Prompt**

You are helping a user generate a logically coherent and fluent paragraph based on retrieved sources related to their writing prompt.
**Below are the relevant sources:** [CONTEXT_BLOCK]
**User's prompt:** [USER_PROMPT]
**Your task:**
- Write a coherent paragraph that responds to the user's prompt.
- Integrate insights **ONLY** from the sources where relevant.
- Add elaborative or connecting content to ensure clarity.
- When referring to any content from the sources, add a tag like [refs:chunk_id1,chunk_id2,...] at the end of the sentence using those sources.
- Output only the paragraph with [refs:...] tags. Do not explain or format anything.

---

We conducted Welch's independent samples t-tests to compare the *Baseline* and the *Personalization* condition across the nine rubric dimensions. Overall, only *Precision* showed a statistically significant difference between conditions, whereas all other dimensions did not reach significance. More specifically, there were no significant differences between the two groups on *Thesis* ($t(38.834) = -0.793$, $p = 0.433$, $d = -0.235$), *Claims* ($t(33.126) = -1.683$, $p = 0.102$, $d = -0.502$), *Evidence for Claims* ($t(40.117) = -1.342$, $p = 0.187$, $d = -0.398$), *Reasoning* ($t(39.976) = -1.405$, $p = 0.168$, $d = -0.417$), *Organization* ($t(40.878) = 0.084$, $p = 0.933$, $d = 0.025$), *Rebuttal* ($t(43.440) = -0.298$, $p = 0.767$, $d = -0.088$), *Conventions* ($t(36.654) = -0.639$, $p = 0.527$, $d = -0.187$), and *Fluency* ($t(43.707) = -0.430$, $p = 0.669$, $d = -0.127$). Only the *Precision* dimension demonstrated a significant difference, with the *Personalization* condition scoring higher than the *Baseline* condition ($t(41.523) = -3.709$, $p < 0.001$, $d = -1.099$), meaning that *Personalization* participants perform better in specific words selection and expressing meanings.